\documentclass[12pt]{article}  

\usepackage{amssymb}
\usepackage{amsmath,bm}

\newtheorem{theorem}{\sc Theorem}[section]

\newtheorem{lemma}[theorem]{\sc Lemma}

\newcommand{\proof}{{\noindent\em Proof. }}

\newcommand{\eps}{\varepsilon}

\newcommand{\proofend}{{\medskip\medskip}}

\title{\LARGE \bf
Inertial Hegselmann-Krause Systems
}

\author{
{\sc Bernard Chazelle}
\thanks{Department of Computer Science,
       Princeton University, 
{\tt chazelle}@{\tt cs.princeton.edu }. This author was supported in part by NSF grants CCF-0963825 and CCF-1420112}%
\and 
{\sc Chu Wang}
\thanks{Program in Applied and Computational Mathematics,
       Princeton University, 
{\tt   chuw}@{\tt math. princeton.edu }}%
}

\begin{document}

\maketitle

\begin{abstract}

We derive an energy bound for {\em inertial Hegselmann-Krause (HK) systems}, which we define as a variant of the classic {\em HK} model in which the agents can change their weights arbitrarily at each step.
We use the bound to prove the convergence of {\em HK} systems with closed-minded agents,
which settles a conjecture of long standing.
This paper also introduces 
{\em anchored HK systems} and show their equivalence
to the symmetric heterogeneous model.
\end{abstract}


\section{Introduction}\label{introduction}

The Hegselmann-Krause model of multiagent consensus
has emerged as a ``model organism" for opinion dynamics~\cite{hegselmanK}.
In an {\em HK} system, 
a collection of $n$ agents, each one represented by a point in ${\mathbb R}^d$,
evolves by applying the following rule at discrete times: move each
agent to the mass center of all the agents within unit distance.
It has been shown that the system always freezes eventually~\cite{chazelle-total, HendrickxB, lorenz05, Moreau2005, nedicT}.
While the model has been the subject of numerous 
studies~\cite{blondelHT07, blondelHT09, castellano2009, krause00, kurzR, Lorenz07}
and much is known about its convergence rate, its {\em heterogeneous} variant remains a mystery \cite{Saber2004,Lorenz2010,Mirtabatabaei2015,Etesami2015,fu2014opinion}.
In that model, each agent can choose its own radius of confidence.  In the {\em HK} model
with {\em closed-minded} agents, all of the agents have radius either 1 or 0. 
While extensive simulations have pointed to the convergence of that system~\cite{Lorenz07,lorenz10, mirtaB, fu2014opinion},
a proof has remained elusive.  This open question has been 
described by a leading researcher as one of the
outstanding gaps in our understanding of opinion dynamics~\cite{Fagnani11}.   
This paper resolves this issue by settling the conjecture in the affirmative: 
{\em HK} systems with closed-minded agents always converge. Our proof entails making
the problem a special case of a much broader class of dynamical systems,
the {\em inertial HK systems} (more on which below).

The relaxation time of the original {\em HK} model
has been shown to be $n^{O(n)}$ in any fixed dimension~\cite{chazelle-total},
a bound later improved to a polynomial bound in both $n$ and $d$~\cite{BhattBCN}.
For the particular case $d=1$, a bound of $O(n^5)$ was established in~\cite{martinezBCF07},
which was lowered to $O(n^4)$ in~\cite{touriN11} and then to~$O(n^3)$ in~\cite{BhattBCN}.
The model can be generalized in various ways, its ultimate expression being
the grand unified model of {\em influence systems}~\cite{chazInfl12},
in which each agent gets to pick its neighbors by following its own distinct, arbitrary criteria.
Oddly, even the most seemingly innocuous modifications of the original {\em HK} model 
have stumped researchers in the field. This is the case of {\em HK} systems with closed-minded
agents, where any agent's radius of confidence is either 0 or 1.
To prove that these systems always converge,
we introduce the more general {\em inertial HK systems}
and establish a bound on their kinetic 2-energy.
We also introduce the {\em anchored} variant of {\em HK} systems
and prove that it is equivalent to the symmetric heterogeneous model.
This fairly surprising result sheds new light
on the convergence properties of these systems.

\subsection{Inertial {\em HK} systems}

Instead of being required to move to the mass center
of its neighbors at each step, each agent
of an {\em inertial HK} system may move toward it by any fraction of length; setting this
fraction to zero makes the agent closed-minded, which means 
that it remains frozen in place. Formally,
the system consists of $n$ agents represented by points
$x_1(t),\ldots, x_n(t)$ in ${\mathbb R}^d$ at time $t=0,1,2,$ etc.
Two agents $i$ and $j$ are said to be {\em neighbors}
if they are within unit distance: $\|x_i(t)- x_j(t)\|_2 \leq 1$.
When the time $t$ is understood, the neighbors
of $i$ form a set we denote by $N_i$; 
these sets form an undirected communication network $G_t$ with 
a self-loop at each of the $n$ nodes.
The dynamics of the system is specified by
\begin{equation}\label{defnX}
x_i(t+1)= (1-\lambda)x_i(t)+ \frac{\lambda}{|N_i|}\sum_{j\in N_i} x_j(t),
\end{equation}
where $\lambda\in [0,1]$ is called the {\em inertia}.
Not only $\lambda$ need not have the same value for all the agents, but
it can be reset to a different value with each application of~(\ref{defnX}).
In this way, we can select any agent to be closed-minded by setting their inertia to $0$.
We can also retrieve the original {\em HK} model by turning all
the inertias to 1. In its full generality, however, an inertial {\em HK} system is
not deterministic. We tackle the issue of convergence 
by turning our attention to their {\em kinetic $s$-energy}. The concept was introduced 
in~\cite{chazelle-total} as a generating function for studying averaging processes
in dynamic networks. It is defined as follows:
$$
K(s)= \sum_{t\geq 0}\,\, \sum_{i=1}^n \|x_i(t+1)-x_i(t)\|_2^s.$$
We provide an upper bound for the case $s=2$.

\medskip
\begin{theorem}\label{2-energyUB}
$\!\!\! .\,\,$
The kinetic $s$-energy of an $n$-agent inertial HK-system whose inertias 
are uniformly bounded from above by $\lambda_0$ satisfies
$K(2)\leq \lambda_0 n^2/4$.
\end{theorem}
\medskip

We use this result to establish the convergence of {\em HK} systems with closed-minded agents.
Note that the convergence is asymptotic. This is even true for $n=2$ with a single
closed-minded agent. Indeed, if the mobile agent is initialized 
close enough to the closed-minded one,
it will eventually converge to it by halving its distance at each step.
The network $G_t$ becomes fixed in this case.
In general, it changes with time, however.
Interestingly, fixed-point attraction does not automatically imply the convergence of
the communication network, so we address this issue separately.

\medskip
\begin{theorem}\label{convergeclosed-minded}
$\!\!\! .\,\,$
An HK system with any number of closed-minded agents converges asymptotically to a fixed-point configuration. The communication network converges for all initial conditions if $d=1$ and for all initial conditions outside
a set of measure zero if $d>1$.
\end{theorem}
\medskip

The specific meaning of this last clause is that, in dimension two and higher,
as long as we perturb the closed-minded agents
by an arbitrarily small amount at the beginning, the communication network $G_t$
will settle to a fixed graph in finite time almost surely.    
The perturbation is likely an unnecessary artifact of the proof and it would be nice to
settle this point. The main open problem, however, is to derive an effective upper bound on 
the relaxation time.

\subsection{Anchored {\em HK} systems}

An anchored  {\it HK}  system consists of $n$ agents, each one represented by a vector $z_k=(x_k(t), y_k)$. The vector is a combination of a mobile part $x_k(t)\in {\mathbb R}^d$
and a static part $y_k\in {\mathbb R}^{d'}$; the dimensions $d$ and $d'$ are the same
for all the agents. 
Two agents $i$ and $j$ are neighbors if and only if $\|z_i(t)-z_j(t)\|_2\le r$, where $r$ is a fixed positive constant. At each step, the mobile part of an agent moves to the mass center of all its neighbors while its anchored part remains fixed. (Note that the averaging
is done one coordinate at a time, so the static coordinates
affect only the neighborhood relationships and do not participate in the averaging itself.)
Anchored  {\it HK} systems capture
a notion of partial closed-mindedness: agents are closed-minded in some coordinates but open-minded in others. Both mobile and anchored parts, on the other hand, affect the communication network.

By contrast, a {\em symmetric heterogeneous HK system} consists of $n$ agents, 
each one represented by a vector $x_k(t)$.
For each pair of agents $(i,j)$, a threshold $r_{ij}$ 
specifies that agents $i$ and $j$ are neighbors at time $t$ 
whenever $\|x_i(t)-x_j(t)\|_2\le r_{ij}$. It is required that $r_{ij}=r_{ji}$
and $r_{ii}\geq 0$ (the latter to create self-loops). 
Note that $r_{ij}=0$ means that $i$ and $j$ are neighbors only when their positions coincide,
while $r_{ij}<0$ implies that $i$ and $j$ are never joined together.
Surprisingly, anchored and symmetric heterogeneous systems are conjugate: 
in other words, there exists a bijection between them that respects their dynamics
and establishes the equivalence of the two systems.
Specifically, we prove the following:

\begin{theorem}\label{anchor&heter}
Given any anchored HK system $z_k(t)=(x_k(t),y_k)$
in ${\mathbb R}^{d}\times {\mathbb R}^{d'}$,
there exists a conjugate symmetric heterogeneous HK system $x'_k(t)$
in  ${\mathbb R}^{d}$.
Conversely, a symmetric heterogeneous HK system of $n$ agents
in  ${\mathbb R}^{d}$ is conjugate
to an anchored HK system $z_k(t)=(x_k(t),y_k)$
with agents in ${\mathbb R}^{d}\times {\mathbb R}^{n-1}$.
In both cases, the conjugacy is formed by the trivial
correspondence: $x_k(t)=x'_k(t)$ for any $k$ and $t$.
Both anchored and symmetric heterogeneous HK systems
converge asymptotically to a fixed configuration.
If there is no pair of agents $(i,j)$ such that $\|y_i-y_j\|_2=r$ in an anchored 
HK system or such that $r_{ij}=0$ in a symmetric heterogeneous HK system, 
then the communication network converges to a fixed graph.
\end{theorem}

While the convergence of symmetric heterogeneous {\em HK}
systems can be inferred directly from known results,
the convergence of the communication networks requires 
special treatment, however.  
An interesting corollary of these results is the convergence of
{\em HK} systems embedded within 
a social network~\cite{das2014,Christakis2008,Weisbuch2004}.
Imagine that the existence of an edge between two agents $i,j$ is
a function not only of their relative distance but also of
a predetermined, fixed relationship. By setting $r_{ij}<0$, we can
enforce the absence of an edge. In this way we can restrict the
{\em HK} action to the edges of a fixed, arbitrary social network,
and still assert convergence.

\section{Inertial {\em HK} Systems \label{sec:inertial}}

The purpose of this section is to prove Theorem~\ref{2-energyUB}.
To do that, we assign each agent $i$ a certain amount of ``money" $C_i(0)$ at the beginning ($t=0$)
and specify a protocol for spending and exchanging it with other agents as
time progresses. If we knew ahead of time the total contribution of agent $i$
to the kinetic 2-energy, we could simply set $C_i(0)$ to that amount and let
the agent ``pay" for its contribution from its own pocket.
This information is not available, however, so we take an initial guess and
set up an exchange protocol so that no agent runs out of money. 
By giving money to their neighbors in a judicious manner, we show how
each agent remains in a position to pay for its share of the 2-energy at each step.
The proof is algorithmic: it is a message-passing protocol that simulates the
update of a distributed Lyapunov function. Our initial guess is
$$C_i(0)=  \sum_{j=1}^n  \min\Bigl\{\, \| x_i(0)-x_j(0)\|_2^2 , \, 1\, \Bigr\}.$$
To specify the exchange protocol, we first simplify the notation as follows:
\begin{equation*}
\begin{cases}
\hspace{.15cm}
\Delta_i= x_i(t+1)-x_i(t) \\
\hspace{.15cm}
d_{ij}= x_i(t)-x_j(t) \\
\hspace{.15cm}
d_{ij}'= x_i(t+1)-x_j(t+1).
\end{cases}
\end{equation*}
The two rules below are applied to every agent $i$ at any time step $t\geq 0$:

\begin{itemize}
\item
For every neighbor of $j$ at time $t$ (which includes $i$ itself),
agent $i$ spends $\|\Delta_i+\Delta_j\|_2^2$ units of money and gives 
to agent $j$ an amount equal to $2(d_{ij}-\Delta_j)^T\Delta_j$.
\item
If agent $j$ becomes a new neighbor of $i$ at time $t+1$ or, conversely,
ceases to be one, then agent $i$ spends $| \|d_{ij}'\|_2^2-1|$.
\end{itemize}
Let $C_i(t)$ be the amount of money held by agent $i$ at time $t$,
and let $N_i^{in}$ (resp. $N_i^{out}$) denote the set of agents 
that are neighbors of $i$ at time $t+1$ (resp. $t$) but not at time $t$ (resp. $t+1$).
Using the symmetry of the neighbor relation, we express the cash flow at time $t$ by
\begin{eqnarray*}
C_i(t+1)-C_i(t) 
&=&
2\sum_{j\in N_i} (d_{ji}- \Delta_i)^T\Delta_i -
2\sum_{j\in N_i} (d_{ij}- \Delta_j)^T\Delta_j 
 \\
&-& \sum_{j\in N_i} \|\Delta_i+\Delta_j\|_2^2 \,\,\, -
\sum_{j\in N_i^{in} \cup N_i^{out}}   | \|d_{ij}'\|_2^2-1|.
\end{eqnarray*}
Since
$
(d_{ji}- \Delta_i)^T\Delta_i - (d_{ij}- \Delta_j)^T\Delta_j 
= d_{ij}^T(\Delta_i- \Delta_j)- 2d_{ij}^T\Delta_i+  \|\Delta_j\|_2^2-  \|\Delta_i\|_2^2$
and, by~(\ref{defnX}),
$\lambda \sum_{j\in N_i}d_{ij} = - |N_i|\Delta_i$,
we have
\begin{eqnarray*}
&&C_i(t+1)-C_i(t) \\
&=& \sum_{j\in N_i} \Bigl\{ 2d_{ij}^T(\Delta_i - \Delta_j) 
    +  \|\Delta_i- \Delta_j\|_2^2 -4d_{ij}^T \Delta_i\Bigr\}  
 - 4 |N_i| \|\Delta_i\|_2^2
- \sum_{j\in N_i^{in} \cup N_i^{out}}   | \|d_{ij}'\|_2^2-1| \\
&=&
\sum_{j\in N_i} \Bigl\{ 2d_{ij}^T(\Delta_i - \Delta_j) +  \|\Delta_i- \Delta_j\|_2^2 \Bigr\}  
+ 4 \bigl(\lambda^{-1}-1\bigr)|N_i| \|\Delta_i\|_2^2
- \sum_{j\in N_i^{in} \cup N_i^{out}}   | \|d_{ij}'\|_2^2-1|.
\end{eqnarray*}
Note that $\lambda=0$ implies that $\Delta_i=0$, so it is understood that
$(\lambda^{-1}-1)|N_i| \|\Delta_i\|_2^2=0$ in the identity above.
Since $d_{ij}' =  d_{ij} + \Delta_i-\Delta_j$, the first summand in the last equality
above is equal to
$\|d_{ij}'\|_2^2 - \| d_{ij}\|_2^2$; therefore
\begin{eqnarray*}
&&C_i(t+1)-C_i(t) \\
&=&
\sum_{j\in N_i} \Bigl\{ \|d_{ij}'\|_2^2 - \| d_{ij}\|_2^2 \Bigr\}  
- \sum_{j\in N_i^{in} \cup N_i^{out}} | \|d_{ij}'\|_2^2-1| 
 + 4 \bigl(\lambda^{-1}-1\bigr)|N_i| \|\Delta_i\|_2^2 \\
&=&
\sum_{j=1}^n \min \bigl\{\, \|d_{ij}'\|_2^2, 1 \,\bigr\} -
\sum_{j=1}^n \min \bigl\{\, \|d_{ij}\|_2^2, 1 \, \bigr\} 
+ 4 \bigl(\lambda^{-1}-1\bigr)|N_i| \|\Delta_i\|_2^2.
\end{eqnarray*}
Since $|N_i|>0$ and $\lambda\le\lambda_0$, it follows that
\begin{eqnarray*}
C_i(t) &\geq& \sum_{j=1}^n \min \bigl\{\, \|d_{ij}\|_2^2, 1 \,\bigr\} + 4 \bigl(\lambda_0^{-1}-1\bigr)\sum_{k=0}^{t-1} \|x_i(k+1)-x_i(k)\|_2^2.
\end{eqnarray*}
Being its own neighbor, agent $i$ spends at least $4\|\Delta_i\|_2^2$ money
at each step. Summing up over all the agents, this amounts to $4K(2)$.
This shows that the initial injection of money allows the system to spend $4K(2)$
and still be left with  $4 \bigl(\lambda_0^{-1}-1\bigr)K(2)$.
Theorem~\ref{2-energyUB} follows from the fact that 
the initial injection of money is at most $n^2$.
\hfill $\Box$
\proofend

\section{{\em HK} Systems with Closed-Minded Agents\label{sec:static}}

This section proves Theorem~\ref{convergeclosed-minded}.
The bound on the kinetic 2-energy shows that the system eventually slows
down to a crawl but it falls short of proving convergence. 
Indeed, an agent moving along a circle by $1/t$ at time $t$ 
contributes finitely to the kinetic 2-energy yet travels an infinite distance.
We prove that {\em HK} systems with closed-minded agents always converge asymptotically.
We treat the one-dimensional separately for two reasons: the proof is entirely self-contained
and the convergence of the communication network does not require perturbation.
In dimension two and higher, we prove that the agents always converge to a fixed position:
the system has a fixed-point attractor.  We show how a tiny random perturbation
ensures that the network eventually settles on a fixed graph.

\subsection{The one-dimensional case}

We begin with the one-dimensional case, which is particularly simple.
By Lemma~\ref{2-energyUB}, we can choose a small enough $\eps>0$ and
an integer $t_\eps$ large enough so that no agent moves by a distance
of more than $\eps$ at any time $t\geq t_\eps$.
Fix $t>t_\eps$ and let
$x_i$ (resp. $N_i$) denote the position (resp. neighbors)
of agent $i$ at time $t$; we use primes 
and double primes to indicate the equivalent quantities for time $t+1$ and $t+2$.
The symmetric difference between $N_i$ and $N_i'$, if nonempty,
is the disjoint union of a set $L_i$ of agents located at $x_i-1\pm O(\eps)$ at times
$t$ and $t+1$ and a set $R_i$ at locations $x_i+1\pm O(\eps)$.
For each subset, we distinguish between the agents of $N_i$ not in $N_i'$
and vice-versa, which gives the disjoint partitions 
$L_i= L_i^{in}\cup L_i^{out}$ and $R_i= R_i^{in}\cup R_i^{out}$. 
The locations $x_i'$ and $x_i''$ of agent $i$ at times $t+1$ and $t+2$ are given by
\begin{equation*}
\begin{cases}
\hspace{.15cm}
|N_i|x_i'= (\,\sum_{j\in N_i\cap N_i'} x_j\,) + (\,\sum_{j\in L_i^{out}\cup R_i^{out}} x_j \,)\\
\hspace{.15cm}
|N_i'|x_i''= (\,\sum_{j\in N_i\cap N_i'} x_j'\,) + (\,\sum_{j\in L_i^{in}\cup R_i^{in}} x_j' \,).
\end{cases}
\end{equation*}
All $x_k'$ and $x_k''$ are of the form $x_k\pm O(\eps)$, so
subtracting the two identities shows that
\begin{eqnarray*}
(|N_i'|- |N_i|)x_i &=&  (|L_i^{in}|- |L_i^{out}|)(x_i-1) \\&+& (|R_i^{in}|- |R_i^{out}|)(x_i+1)  \pm O(\eps n).
\end{eqnarray*}
Since the dynamics is translation-invariant, we can assume that $x_i=0$. Setting $\eps$ small enough,
the integrality of the set cardinalities implies that the net flow of neighbors on the left of agent $i$
is the same as it is on the right:
\begin{equation}\label{InAndOut}
|L_i^{out}|- |L_i^{in}| =|R_i^{out}|- |R_i^{in}|.
\end{equation}
Among all the agents undergoing a change of neighbors between times $t$ and $t+1$,
pick the one that ends up the furthest to the right at time $t+1$, choosing
the one of largest index $i$ to break ties.
We distinguish between two cases:
\begin{enumerate}
\item
$x_i'\geq x_i$: No agent of $R_i^{out}$ can be closed-minded;
nor can it be mobile since, ranks being preserved, it would provide
an agent undergoing a change of neighbors and landing to the right of $i$ at time $t+1$,
in contradiction with the definition of $i$. It follows that $R_i^{out}$ is empty,
which in turn implies that $L_i^{in}$ is not, since by our choice of $i$
not all four terms in~(\ref{InAndOut}) can be zero. Since agent $i$ is
not moving left, neither is any agent $j$ of $L_i^{in}$.
Its set $N_j$ of neighbors changes
between time $t$ and $t+1$ and $R_j^{out}$ is empty.
To see why the latter is true, we first note that $N_j$
cannot lose any closed-minded agent to the right. Also, since any mobile agent
in $R_j^{out}$ is to the left of $i$ at time $t$, it
stays to the left of it by conservation of ranks;
hence the agent remains a neighbor of $j$, a contradiction.
The argument so far uses the rightmost status of agent $i$ only to assert
that $R_i^{out}$ is empty. This means we are back to square one
and we can proceed inductively, eventually reaching a contradiction.
\item
$x_i' < x_i$: The key observation is that our previous argument never uses
time directionality, so we can exchange the role of $t$ and $t+1$, which implies
that now $x_i' > x_i$. Note that the superscripts {\em in} and {\em out} must be swapped.
While we chose $i$ as the mobile agent landing furthest to the right, by symmetry
we must now choose the one starting the furthest to the right: of course, since mobile
agents can never cross this make no difference.
\end{enumerate}

We conclude that each agent is now endowed with a fixed set of neighbors, so the
dynamics is specified by the powers of a fixed stochastic matrix with positive diagonal, 
which are well known to converge. The system is attracted to a fixed point at 
an exponential rate, but of course we have no a priori bound on the time it takes
to fall into that basin of attraction. The communication network converges.

\subsection{The higher-dimensional case\label{3.2}}

Generalizing the previous argument to higher dimension
fails on several counts, the most serious one being the
loss of any left-right ordering.   
We follow a different tack, which
begins with a distinction between two types of agents. 
An agent is {\em trapped} at time $t$ if there exists a path 
in the current communication graph leading to a closed-minded agent; 
it is said to be {\em free} otherwise. There exists a time $t_o$
after which the agents fall into two categories: 
some of them are never trapped past $t_0$ and are called {\em eternally free};
the others are {\em chronically trapped} (ie, trapped an infinite number of times).
As we did before,
we pick a parameter $\eps>0$ (to be specified below) and
$t_\eps>t_o$ large enough so that no agent moves by a distance
of more than $\eps$ at any time $t\geq t_\eps$. If two agents ever get
to share the same position, their fates become completely tangled since they can never
again get separated. Since such merges occur fewer than $n$ times, we can make
$t_\eps$ big enough, if necessary, so that all merges are in the past.
To summarize, past $t_\eps$, the mobile agents move by increments less than $\eps$,
no merging occurs, and the system consists only of eternally free and chronically trapped agents.

At any time, the state system is represented by a $n$-by-$d$ matrix whose $i$-th row encode
the position of agent $i$ in ${\mathbb R}^d$. The matrix consists of two parts:
$x$ for the mobile agents and $y$ for the closed-minded ones. A transition of the system
is a linear map of the form $x\mapsto Ax+By$, where each row of the nonnegative 
matrix $(A\,|\,B)$ sums up to~1.

\begin{lemma}\label{freeStuck}
$\!\!\! .\,\,$
Past $t_\eps$, no agent can move while free.  
\end{lemma}

\proof
Fix $t\geq t_\eps$ and consider a connected component $\mathcal C$ of the graph induced by the free agents.
If $z$ denotes its position matrix at time $t$ and $k$ its number of rows, then
$z'= Cz$, where primes refer to time $t+1$ and
$C$ is a $k$-by-$k$ stochastic matrix for a random walk in the
undirected graph $\mathcal C$. Because the graph is connected, the eigenvalue~1
of $C$ is simple, so the null space of $I-C$, and hence of $(I-C)^T(I-C)$, is spanned
by ${\mathbf 1}$. By Courant-Fischer, therefore, any vector $u$ normal to ${\mathbf 1}$ satisfies
$\| (I-C)u\|_2\geq \sigma \|u\|_2$, where $\sigma$ is the smallest positive singular value of $I-C$.
If we define $\bar z= z- \frac{1}{k}{\mathbf 1}{\mathbf 1}^Tz$, it immediately follows 
that
$$
\sigma \|\bar z\|_2\leq \| (I-C)\bar z\|_2 = \|(I-C)z\|_2= \|z-z'\|_2\leq \eps \sqrt{n}.
$$
Setting $\eps< \frac{1}{2} \sigma/\sqrt{n}$ ensures that any two of the $k$ agents are within unit distance.
It follows that $\mathcal C$ is the complete graph and $C= \frac{1}{k}{\mathbf 1}{\mathbf 1}^T$.
Since the agents can no longer merge, the only option left is for all $k$ of them to be
already merged at time $t$, hence unable to move.
\hfill $\Box$
\proofend

The lemma implies that eternally free agents can never move again past $t_\eps$.
Indeed, it shows that an eternally free agent can only move if it is joined
to a trapped one, which, by definition, it cannot be. Since eternal freedom keeps
the agents from playing any role after time $t_\eps$, we might as well assume
that all the mobile agents in the system are chronically trapped. This means that, at all instants,
either an agent is trapped (ie, joined to a closed-minded agent via a path) or it is 
{\em isolated}, meaning that
the other agents are either merged with it or at distance
greater than one. An agent cannot move while isolated.

The position matrix $z$ of the $k$ trapped agents
at time $t\geq t_\eps$ satisfies the relation $z'= Tz+ Uy$, where primes denote 
time $t+1$ and the $k$-by-$n$ matrix $(T\,|\,U)$ has each row summing up to 1.
Being trapped implies that $U$ is not the null matrix. In fact,
viewed as a Markov chain, the trapped agents correspond to transient states, which
means that $T^k$ tends to the null matrix as $k$ goes to infinity. This shows that
$T$ cannot have 1 as an eigenvalue; therefore $I-T$ is nonsingular.
Let $\mu$ be a uniform upper bound on the singular values
of all the (so-called fundamental) matrices $(I-T)^{-1}$; since their number is finite, so is $\mu$.
Since $z'= Tz+Uy$ and $\|z'-z\|_2\leq \eps \sqrt{n}$,
the matrix $z$ is very close to $(I-T)^{-1}Uy$; specifically,
\begin{equation}\label{zClose}
\| z- (I-T)^{-1}Uy\|_2=  \|(I-T)^{-1}(z-z')\|_2 \leq \mu\|z-z'\|_2 \leq \mu \eps \sqrt{n}.
\end{equation}
A matrix of the form $(I-T)^{-1}Uy$ is called an {\em anchor}.
Since the set of all possible anchors (for given $y$) is finite,
the minimum (Frobenius-norm) distance $r$ between any two distinct anchors
is strictly positive. The value of $r$ does not
depend on $\eps$, so we can always lower the value of the latter, if necessary, to
ensure that $r>  (1+2\mu) \eps \sqrt{n}$. 

By~(\ref{zClose}) and Lemma~\ref{freeStuck}, we know that, at any time $t$ past
$t_\eps$, any mobile agent is either stuck in place (if free) or at distance at most
$\mu\eps\sqrt{n}$ away from an anchor. As a result, no agent can ever change anchors 
since this would necessitate a one-step leap of at least $r-2\mu\eps\sqrt{n}>\eps\sqrt{n}$
for the positional matrix, 
hence the displacement of an agent by a distance of at least $\eps$, which has been ruled out.
Since the argument holds for any $\eps$ small enough, each mobile agent is thus constrained
to converge toward its chosen anchor.
This concludes the proof that all agents converge to a fixed point in ${\mathbb R}^d$.
The convergence is asymptotic and no bound can be inferred directly from our analysis.

The result does not imply that the communication network
should also converge to a fixed graph. The lack of convergence points to a situation
where the agents are still moving in increasingly small increments, yet edges of the
network keep switching forever.  This can only occur if at least one pair of anchor points
are at distance 1:  by anchor point, we mean the points formed by any row of an anchor matrix
or of $y$.  The key observation is that all the anchor points 
are convex combinations of the rows of $y$, so an interdistance of 1 is expressed by
an equality of the form $\|v^Ty\|_2=1$. There are only a finite set of such equalities
to consider and each one denotes an algebraic surface of codimension 1. 
Any random perturbation of the closed-minded agents will result in the convergence of
the communication network almost surely.
This completes the proof of Theorem~\ref{convergeclosed-minded}.
\hfill $\Box$
\proofend

\section{Anchored and Symmetric Heterogeneous {\it HK} Systems \label{sec:anchor}}

This section proves Theorem \ref{anchor&heter}.
We begin with a proof of the conjugacy between the two 
types of {\em HK} systems. 

\subsection{The bijection relation}

To express an anchored {\it HK} system $z(t)=(x_k(t),y_k)$ as
a symmetric heterogeneous one is straightforward.
We have the equivalence 
\begin{equation}\label{eq:eqi}
\|z_i(t)-z_j(t)\|_2^2 \le r^2 \Leftrightarrow
\|x_i(t)-x_j(t)\|_2^2 \le r^2- \|y_i-y_j\|_2^2. 
\end{equation}
We define $r_{ij}=\sqrt{r^2- \|y_i-y_j\|_2^2}$ if the right hand side of \eqref{eq:eqi} is non-negative, and $r_{ij}=-1$ otherwise. 
Then the system $x_k(t)$ together with thresholds $r_{ij}$ forms
a symmetric heterogeneous {\it HK} system.
Notice that the equivalence \eqref{eq:eqi} ensures that the communication graphs of the given anchored {\it HK} system and its corresponding symmetric 
heterogeneous {\it HK} counterpart are identical. 

For the other direction, 
we need to lift the given symmetric heterogeneous {\it HK} system to an anchored {\it HK}
version.  We need the following lemma, whose proof can be found in the Appendix.

\begin{lemma}\label{lemma:anchor}
$\!\!\! .\,\,$
For any $n$-by-$n$ symmetric matrix $R=(r_{ij})$ with no negative terms
in the diagonal, there exist $r>0$ and vectors $y_k\in\mathbb{R}^{n-1}$
$(1\leq k\leq n)$, such that
\begin{equation}
\|y_i-y_j\|_2=\sqrt{r^2-r_{ij}^2\text{~sign}\,(r_{ij})},
\end{equation}
for any $i\neq j$; here $\text{sign}\,(x)=1$ if $x\geq 0$ and $-1$ otherwise.
\end{lemma}

Given a symmetric heterogeneous {\em HK} system $x_k(t)$, 
we choose the anchors $y_k$ by appealing to Lemma \ref{lemma:anchor}.
For any $r_{ij}\ge 0$, it then follows that 
\begin{equation}
\|x_i(t)-x_j(t)\|_2^2\le r_{ij}^2 \Leftrightarrow
\|x_i(t)-x_j(t)\|_2^2 + \|y_i-y_j\|_2^2\le r^2,
\end{equation}
and for any $r_{ij}< 0$, and we always have
\begin{equation}
\|x_i(t)-x_j(t)\|_2^2 + \|y_i-y_j\|_2^2>r^2,
\end{equation}
for any $i\neq j$, which prevents any edge between $i$ and $j$.
This means that the dynamics of the symmetric heterogenous {\em HK} system
coincides precisely with that of the mobile part of the lifted anchored system.

\bigskip
\noindent
{\em Remark:}
Lemma \ref{lemma:anchor} asserts that, given $(n-1)n/2$ lengths $d_{ij}$ $(i\neq j)$
of the form $( r^2- r_{ij}^2\text{~sign}(r_{ij}) )^{1/2}$,
we can find $n$ points $y_k\in \mathbb{R}^{n-1}$
such that the pairwise distance $\|y_i-y_j\|_2=d_{ij}$.
Notice that, if $d_{ij}$ itself is arbitrary, this is not always possible.
For example, in the case $n=3$, the problem is equivalent 
to finding a triangle in $\mathbb{R}^{2}$ with each side length given.
The problem is solvable if and only if the three lengths satisfy the triangle inequality.
In our case, however, there is an extra parameter $r$ that we can use.
Intuitively, if we choose a large $r$ such that all the $|r_{ij}|$ are relatively small,
then the problem of finding $y_k$ is equivalent to finding an almost regular polytope, each edge of which is roughly of the same length~$r$.

\subsection{Proof of convergence}

The fixed-point attraction of symmetric heterogeneous {\em HK}
systems can be inferred directly from
known results about infinite products of type-symmetric 
stochastic matrices~\cite{chazelle-total, HendrickxB, lorenz05, Moreau2005}.
The same holds of anchored systems.
In both cases, given any $\eps>0$ and any 
initial condition, the $n$ agents will eventually reach a ball of radius $\eps$
that they will never leave; we call this
{\em $\eps$-convergence}. We study the conditions for this to imply that
the corresponding communication networks themselves converge
to a fixed graph. It suffices to consider the case of a symmetric
heterogeneous {\em HK} system. 
Consider a connected component $\mathcal{C}$ of the graph and let $z$ and $z'=Cz$ denote the corresponding position matrices at time $t$ and $t+1$, where $C$ is 
the corresponding $k$-by-$k$ stochastic matrix associated with $\mathcal{C}$. 
As we did in the proof of Lemma~\ref{freeStuck}, we define $\sigma$ to
be a uniform lower bound on any positive singular value of $I-C$ for any
such matrix $C$. Setting 
$$\eps= \frac{\sigma}{2\sqrt{n}}\, \min_{r_{ij}>0}r_{ij}$$
implies that
$$\|\bar{z}\|_2 \le \frac{1}{\sigma} \|(I-C)\bar{z}\|_2 = \frac{1}{\sigma} \|z-z'\|_2 \le \frac{\sqrt{n}\epsilon}{\sigma}\le\frac{1}{2} \min_{r_{ij}>0}r_{ij} , $$
where
$\bar z= z- \frac{1}{k}{\mathbf 1}{\mathbf 1}^Tz$ is the
projection of $z$ onto the orthogonal space of ${\mathbf 1}$.
It follows that, 
for any pair $(i,j)$ in $\mathcal{C}$ such that $r_{ij}>0$, 
there will be an edge between $i$ and $j$. With the assumption $r_{ij}\neq0$,
the communication graph is now fixed
and convergence proceeds at an exponential rate
from that point on. The bijection result of the previous section
shows that the condition $r_{ij}=0$ corresponds to 
$\|y_i-y_j\|_2=r$ in the case of anchored systems.
This concludes the proof of Theorem~\ref{anchor&heter}.
\hfill$\Box$

\vspace{1cm}


\newpage

\section*{Appendix}

Our proof of Lemma \ref{lemma:anchor} relies on two technical facts.
For convenience, we use bold letters to denote vectors;
for example, $u_{k}$ denotes the $k$-th coordinate
of vector $\bm{u}$.

\paragraph{Fact A.} \
{\em There exist $n+1$ vectors $\bm{u}^{(k)}\in\mathbb{R}^{n}$ $(0\le k \le n)$
such that $\|\bm{u}^{(i)}-\bm{u}^{(j)}\|_2=1$
$(0\leq i<j\leq n)$, $u_{i}^{(k)}=0$ for $i > k\geq 0$ and 
all $u_k^{(k)}$ exceed $1/\sqrt{2}$ and decrease as $k$ grows.}

\medskip
\begin{proof}
Proceeding by induction, we write
$\bm{u}^{(0)}=\bm{0}$, $\bm{u}^{(1)}=\bm{e}_1$ and $\bm{u}^{(2)}=\frac{1}{2}\bm{e}_1+\frac{\sqrt{3}}{2}\bm{e}_2$,
where $\bm{e}_i$ is the unit vector in the $i$-th dimension.
Assume we already constructed $\bm{u}^{(k)}$ ($0\leq k\leq m<n$)
such that 
$u_{i}^{(k)}=0$ for $i>k$ and $u_{k}^{(k)}>1/\sqrt{2}$.
Then we can write $\bm{u}^{(k)}$~as 
$$\bm{u}^{(k)}=\sum_{i=1}^{k}u_{i}^{(k)}\bm{e}_i, ~~~~k=1,2,\ldots, m.$$
We  define 
$$\bm{u}^{(m+1)}=\sum_{i=1}^{m-1}u_{i}^{(m)}\bm{e}_i+
\Big{(}u_m^{(m)}-\frac{1}{2u_m^{(m)}}\Big{)}\bm{e}_{m}+\sqrt{1-\frac{1}{4\big{(}u_m^{(m)}\big{)}^2}}\, \bm{e}_{m+1}.$$
Since $u_{m}^{(m)}> 1/\sqrt{2}$, we have 
$$u_{m+1}^{(m+1)}> \sqrt{1-\frac{1}{4(1/\sqrt{2})^2}}=\frac{1}{\sqrt{2}}.$$
For $k=0,1,\ldots, m,$
$$\|\bm{u}^{(m+1)}-\bm{u}^{(k)}\|_2^2
=\|\bm{u}^{(m)}-\bm{u}^{(k)}\|_2^2
+ u_m^{(k)}/ u_m^{(m)}
= (1-\delta_{km}) + u_m^{(k)}/ u_m^{(m)} =1.$$
Notice that, for $0\leq k<n$,
$$\big{(}u_{k+1}^{(k+1)}\big{)}^2-\big{(}u_{k}^{(k)}\big{)}^2=\bigg{(}1-\frac{1}{4\big{(}u_k^{(k)}\big{)}^2}\bigg{)}-\big{(}u_{k}^{(k)}\big{)}^2
=-\bigg{(}u_k^{(k)}-\frac{1}{2u_k^{(k)}}\bigg{)}^2\le 0,$$
which proves the monotonicity claim.

\end{proof}

\paragraph{Fact B.} \
{\em 
For any integer $n>0$, there is a positive number $\gamma$ depending on $n$
such that, for any $t_{ij}$ satisfying $|1-t_{ij}| \le \gamma$
and $t_{ij}=t_{ji}$ $(0\le i< j\le n)$, there exist vectors $\bm{y}^{(k)}\in\mathbb{R}^{n}$
$(0\le k \le n)$
such that $\|\bm{y}^{(i)}-\bm{y}^{(j)}\|_2=t_{ij}$,
for $0\leq i<j\leq n$.}

\medskip
\begin{proof}
We make repeated use of the matrix infinity norm.
Recall that if $M$ is a $p$-by-$q$ matrix, its infinity norm
is defined as the  maximum absolute row sum of $M$:
$$\|M\|_\infty:=\max_{1\le i\le p} \sum_{j=1}^{q}|m_{ij}|.$$
As one would expect of a matrix norm, the infinity norm is
submultiplicative: $$\|MN\|_\infty\le \|M\|_\infty \|N\|_\infty,$$
for any $p$-by-$q$ matrix $M$ and $q$-by-$r$ matrix $N$.
We define a constant 
$$\alpha=5n+\max_{1\le k\le n}\|C_k^{-1}\|_\infty,$$
where $C_k$ is the 
$k$-by-$k$ matrix 
whose $i$-th row consists of the first  $k$ elements of the vector $\bm{u}^{(i)}$ in Fact A.
Note that $C_k$ is lower-triangular and invertible.
Let $\gamma=\alpha^{-4n}$. The intuition of the
proof is that the vectors $\bm{y}^{(k)}$ we are seeking
should be close to the vectors $\bm{u}^{(k)}$.
We build the desired vectors by induction. 
Let $\bm{y}^{(0)}=\bm{0}$ and $\bm{y}^{(1)}=t_{01}\bm{e}_1$.
Then it is obvious that $\|\bm{y}^{(0)}-\bm{y}^{(1)}\|_2=t_{01}$ and $\bm{y}^{(0)}$ and $\bm{y}^{(1)}$ are close to the vectors from Fact A:
$$\|\bm{y}^{(0)}-\bm{u}^{(0)}\|_\infty=0 < \gamma,$$
$$\|\bm{y}^{(1)}-\bm{u}^{(1)}\|_\infty=|t_{01}-1| \le \gamma \le \alpha^{4}\gamma.$$
Suppose $\bm{y}^{(0)},\bm{y}^{(1)},\dots,\bm{y}^{(k-1)}$ 
have been specified such that $y_i^{(j)}=0$ for $i>j$,
\begin{equation}\label{y-u}
\|\bm{y}^{(i)}-\bm{u}^{(i)}\|_\infty\le \alpha^{4i}\gamma~~~~(0\le i \le k-1),
\end{equation}
and
$$\|\bm{y}^{(i)}-\bm{y}^{(j)}\|_2=t_{ij}~~~~(0\le i< j \le k-1).$$
We need to show is that there exists a vector $\bm{y}^{(k)}$ such that $y_i^{(k)}=0$ for $i>k$,
\begin{equation} \label{toshow2}
\|\bm{y}^{(k)}-\bm{u}^{(k)}\|_\infty\le \alpha^{4k}\gamma
\end{equation}
and
\begin{equation} \label{toshow1}
\|\bm{y}^{(k)}-\bm{y}^{(i)}\|_2 =t_{ik}~~~~(0\le i \le k-1).
\end{equation}
This last relation is equivalent to
\begin{equation}\label{big}
\sum_{j=1}^i  \big{(}y_j^{(k)}-y_j^{(i)}\big{)}^2 +\sum_{j=i+1}^k  \big{(}y_j^{(k)}\big{)}^2=t_{ik}^2~~~~(0\le i \le k-1).
\end{equation}
By subtracting the equations for $1\le i \le k-1$ from the one
for $i=0$, we get a linear system for
$\hat{\bm{y}}:=(y_1^{(k)},y_2^{(k)},\dots,y_{k-1}^{(k)})^T$: 
$$A\hat{\bm{y}}=\bm{b}.$$
Here the $(k-1)\times (k-1)$ matrix $A$ is a lower triangular matrix where $A_{ij}=y_j^{(i)}$
($i\geq j$)
and $\bm{b}$ is a  $(k-1)$ dimensional column vector where 
$$b_i=\frac{1}{2}\Bigl( t_{0k}^2-t_{ik}^2+\sum_{j=1}^{i}\big{(}y_{j}^{(i)}\big{)}^2 \Bigr ).$$
We derive similar relations from Fact A:
\begin{equation}\label{big2}
\sum_{j=1}^i  \big{(}u_j^{(k)}-u_j^{(i)}\big{)}^2 +\sum_{j=i+1}^k  \big{(}u_j^{(k)}\big{)}^2=1~~~~(0\le i \le k-1),
\end{equation}
which implies a linear system for $\hat{\bm{u}}:=(u_1^{(k)},u_2^{(k)},\dots,u_{k-1}^{(k)})^T$: 
$$C\hat{\bm{u}}=\bm{d},$$
where $C$ is shorthand for $C_{k-1}$
and $d_i=\frac{1}{2}\sum_{j=1}^{i}\big{(}u_{j}^{(i)}\big{)}^2$.
We already observed that $C$ is nonsingular; we note that, by~\eqref{y-u} 
and $u_i^{(i)}>1/\sqrt{2}$, the same is true of $A$.
Next, we derive upper bounds on the length of the vector
$\bm{b}$ and its distance from~$\bm{d}$.
By $|1-t_{ij}|\le \gamma$ and $\gamma<1/2$,
\begin{equation}\label{bd1}
|t_{0k}^2-t_{ik}^2|=|t_{0k}+t_{ik}||t_{0k}-t_{ik}|\le (2+2\gamma)\cdot 2\gamma < 6\gamma.
\end{equation}
By our induction hypothesis~\eqref{y-u}, the fact that $|y_{j}^{(i)}|\le 1+\gamma$, and the definition of $\gamma$, we have
\begin{equation}\label{bd2}
\left | \big{(}y_j^{(i)}\big{)}^2-\big{(}u_j^{(i)}\big{)}^2 \right |=|y_j^{(i)}+u_j^{(i)} ||y_j^{(i)}-u_j^{(i)} |\le 
(2+\alpha^{4i}\gamma)\cdot \alpha^{4i}\gamma < 3 \alpha^{4(k-1)}\gamma.
\end{equation}
Thus, by~(\ref{bd1}, \ref{bd2}), 
\begin{equation} \label{e1}
\|\bm{b}-\bm{d}\|_\infty\le 3(1+n\alpha^{4(k-1)}/2)\gamma.
\end{equation} 
By  inequality \eqref{bd1} and the fact that $\gamma$ is small enough, we have
\begin{equation} \label{e2}
\|\bm{b}\|_{\infty}\le \frac{1}{2}
\Bigl(\max _{1\le i\le k }|t_{0k}^2-t_{ik}^2|
+\max _{1\le i\le k }\|\bm{y}^{(i)}\|_2^2 \Bigr)
< \frac{1}{2}(6\gamma+(1+\gamma)^2)<1.
\end{equation}
We also claim that 
\begin{equation}\label{e4}
\|A^{-1}-C^{-1}\|_\infty\le 2n\alpha^{4k-2}\gamma.
\end{equation} 
Here is why. First, 
notice that \eqref{y-u} implies $\|A-C\|_\infty\le n\alpha^{4(k-1)}\gamma$.
Then based on the definition of $\alpha$, we have $\|C^{-1}\|_\infty < \alpha$,
and hence
\begin{equation}\label{e5}
\|C^{-1}(A-C)\|_\infty\le \|C^{-1}\|_\infty\|A-C\|_\infty < n\alpha^{4k-3}\gamma.
\end{equation} 
The right hand side of the above inequality is smaller than $1/2$ based on the definition of $\gamma$,
which allows us to expand the matrix inverse $[I+C^{-1}(A-C)]^{-1}$ as
$$[I+C^{-1}(A-C)]^{-1}=I+\sum_{i=0}^\infty (-1)^i [C^{-1}(A-C)]^i,$$
from which it follows that
\begin{equation}\label{e3}
\|[I+C^{-1}(A-C)]^{-1}\|_\infty\le 2.
\end{equation} 
Notice that
$$A^{-1}-C^{-1}=[I+C^{-1}(A-C)]^{-1}C^{-1}(C-A)C^{-1},$$
then inequality \eqref{e4} directly follows from inequalities \eqref{e5} and \eqref{e3}.
By~(\ref{e1}, \ref{e2}, \ref{e4}) and the fact that $\|C^{-1}\|_{\infty}<\alpha$, finally we have 
\begin{eqnarray*}
\|\hat{\bm{y}}-\hat{\bm{u}}\|_{\infty}
&=&\|A^{-1}\bm{b}-C^{-1}\bm{d}\|_\infty\\
&=&\|(A^{-1}-C^{-1})\bm{b}+C^{-1}(\bm{b}-\bm{d})\|_\infty\\
&\le& \|(A^{-1}-C^{-1})\|_\infty \|\bm{b}\|_\infty+\|C^{-1}\|_\infty\|(\bm{b}-\bm{d})\|_\infty\\
&\le&  2n\alpha^{4k-2}\gamma+ 3(1+ n\alpha^{4(k-1)}/2)\alpha\gamma
< \alpha^{4k-1}\gamma.
\end{eqnarray*}
This shows that 
\begin{equation}\label{induction1}
|y_j^{(k)}-u_j^{(k)}|\le \alpha^{4k-1}\gamma ~~~~(1\le j \le k-1).
\end{equation}
In turn, this implies that
\begin{equation}\label{e6}
\left | \big{(}y_j^{(k)} \big{)}^2 -\big{(}u_j^{(k)} \big{)}^2 \right |=|y_j^{(k)}+u_j^{(k)}||y_j^{(k)}-u_j^{(k)}|
<(2+\alpha^{4k-1}\gamma) \alpha^{4k-1}\gamma <3\alpha^{4k-1}\gamma.
\end{equation}
It suffices now to set the remaining (nonzero) coordinate of $\bm{y}^{(k)}$ yet to be
specified, which is $y_k^{(k)}$.
Recall that it must satisfy
$$
\sum_{j=1}^k  \big{(}y_j^{(k)}\big{)}^2=t_{0,k}^2
$$
and, by our construction, this single equality suffices to imply all of~(\ref{toshow1}). 
This implies a unique setting of (positive) $y_k^{(k)}$, so we need only
be concerned with~(\ref{toshow2}) and the positivity of $\big{(}y_k^{(k)}\big{)}^2$.
Since $|1-t_{0k}^2|=|1-t_{0k}||1+t_{0k}|\le \gamma (2+\gamma) <3 \gamma$,
inequality  \eqref{big2} for $i=0$, combined with 
\eqref{bd2}, establishes that 
$$\left | \big{(}y_k^{(k)}\big{)}^2-\big{(}u_k^{(k)}\big{)}^2 \right |
\le \sum_{i=1}^{k-1}\left | \big{(}y_i^{(k)}\big{)}^2-\big{(}u_i^{(k)}\big{)}^2\right |+|1-t_{0k}^2 |\le 3(1+n\alpha^{4k-1})\gamma.
$$
Since $u_k^{(k)}> 1/ \sqrt{2}$, it follows that
$$\big{(}y_k^{(k)}\big{)}^2>\frac{1}{2}-3(1+n\alpha^{4k-1})\gamma>0.$$
Furthermore,
\begin{equation}\label{induction2}
|y_k^{(k)}-u_k^{(k)}|=\frac{\big{|}\big{(}y_k^{(k)}\big{)}^2-\big{(}u_k^{(k)}\big{)}^2\big{|}}{y_k^{(k)}+u_k^{(k)}}
\le 3\sqrt{2}(1+n\alpha^{4k-1})\gamma<\alpha^{4k}\gamma.
\end{equation}
In conjunction with~\eqref{induction1}, this establishes~\eqref{toshow2},
and completes the inductive construction.
\hfill $\Box$
\end{proof}

It should be noted that Fact B can also be proven
via the implicit function theorem and a perturbation argument based on Fact A.
The benefit of the proof given above is to provide an explicit construction.

\paragraph{Lemma 4.1}\
{\em
For any $n$-by-$n$ symmetric matrix $R=(r_{ij})$ with no negative terms
in the diagonal, there exist $r>0$ and vectors $y_k\in\mathbb{R}^{n-1}$
$(1\leq k\leq n)$, such that
\begin{equation}
\|y_i-y_j\|_2=\sqrt{r^2-r_{ij}^2\text{~sign}\,(r_{ij})},
\end{equation}
for any $i\neq j$; here $\text{sign}\,(x)=1$ if $x\geq 0$ and $-1$ otherwise.
}

\medskip
\begin{proof}
Choose a sufficiently large $r$ such that $$\max_{i,j}|r_{ij}|<\gamma r,$$
where $\gamma$ is the small positive constant from Fact B. 
We set $t_{ij}$ to $\sqrt{1-r^2_{ij}\text{~sign}(r_{ij})/r^2}$ and easily verify that $|1-t_{ij}| \le \gamma$.
Fact B guarantees
the existence of vectors $\bm{z}_k\in\mathbb{R}^{n-1}$ ($1\leq k\leq n$)
such that $\|\bm{z}_i-\bm{z}_j\|_2=t_{ij}$.
Setting $y_k=r\bm{z}_k$ satisfies the requirements.
\hfill $\Box$
\end{proof}

\end{document}